\documentclass[11pt,a4paper]{article}
\usepackage{jheppub}

\newcommand{\be}{\begin{equation}}
\newcommand{\beq}{\begin{equation}}
\newcommand{\ee}{\end{equation}}
\newcommand{\eeq}{\end{equation}}
\newcommand{\ba}{\begin{aligned}}
\newcommand{\ea}{\end{aligned}}
\newcommand{\bea}{\begin{eqnarray}}
\newcommand{\eea}{\end{eqnarray}}
\newcommand{\bear}{\begin{eqnarray}}
\newcommand{\eear}{\end{eqnarray}}
\newcommand{\bean}{\begin{eqnarray*}}
\newcommand{\eean}{\end{eqnarray*}}

\def\ie{\hbox{\it i.e. }}

\newcommand{\reef}[1]{(\ref{#1})}

\title{On Lovelock analogues of the Riemann tensor}

\author[a]{Xi\'an O. Camanho} \emailAdd{xian.camanho@aei.mpg.de}

\author[b,c]{and Naresh Dadhich} \emailAdd{nkd@iucaa.ernet.in}

\affiliation[a]{Max-Planck-Institut f\"ur Gravitationsphysik, Albert-Einstein-Institut, 14476 Golm, Germany}
\affiliation[b]{Centre for Theoretical Physics, Jamia Millia Islamia, New Delhi~110025, India}
\affiliation[c]{Inter-University Centre for Astronomy and Astrophysics, Post Bag 4, Pune 411 007, 
India}

\abstract{It is possible to define an analogue of the Riemann tensor for $N$th order Lovelock
gravity, its characterizing property being that the trace of its Bianchi derivative yields the corresponding analogue of the Einstein tensor. Interestingly there exist two parallel but distinct such analogues and the main purpose of this note is to reconcile both these formulations. In addition we will show that any pure Lovelock vacuum in odd $d=2N+1$ dimensions is Lovelock flat, {\it i.e.} any vacuum solution of the theory has vanishing Lovelock-Riemann tensor. Further, in presence of cosmological constant it is the Lovelock-Weyl tensor that vanishes.}


\preprint{${\rm AEI}-2015-XXX$}

\begin{document}

\maketitle

\allowdisplaybreaks

\section{Introduction} 

In order to write an equation of motion for Einstein gravity, one has to obtain a divergence free second rank symmetric tensor constructed solely from the metric and the Riemann curvature - the Einstein tensor. This is usually done by varying the Einstein-Hilbert action, the scalar curvature $R$,  relative to the metric tensor. Alternatively one can obtain the same result by invoking the differential geometric property that the Bianchi derivative of the Riemann tensor identically vanishes - the Bianchi identity. From the trace of this identity we can then extract the required Einstein tensor. This is a very neat and elegant purely geometric way to get to the equation of motion. The most natural generalization of Einstein gravity in higher dimensions is Lovelock gravity, whose equation of motion inherits the basic property of being second order, though polynomial in curvature. The natural question then arises, could the same geometric method used for Einstein gravity also work for Lovelock theories? The answer is yes. In \cite{dad1} one of the authors defined an $N$th order Lovelock analogue of the Riemann curvature, which is a homogeneous polynomial in the Riemann tensor. Even though this tensor does not satisfy the Bianchi identity, the trace of its Bianchi derivative vanishes yielding the $N$th order Lovelock analogue of the Einstein tensor. This tensor agrees with the one obtained by varying $N$th order Lovelock Lagrangian and is divergence free.  \\

There is an alternative formulation due to Kastor \cite{kast} that also leads to the definition of a different higher order analogue of the Riemann tensor. His construction has a much richer geometric structure as it involves a $4N$th rank tensor as its basic object. This higher rank tensor does satisfy the Bianchi identity, {\it i.e.} its Bianchi derivative vanishes, and again the trace of this identity leads to the Einstein analogue. Interestingly the Einstein analogues obtained in Dadhich's and Kastor's formulations agree and therefore lead to the same equation of motion, both descriptions are dynamically equivalent. This had to be the case as the Lovelock Lagrangian is unique at each order. The main aim of this note is to reconcile these two parallel formulations and also to illuminate a universal property of pure Lovelock gravity that distinguishes between odd $d=2N+1$ (the critical dimension) and even $2N+2$ (or higher) dimensions. Dadhich {\it et al.} \cite{dgj} considered pure Lovelock static vacuum solutions and established that pure Lovelock gravity in odd $d=2N+1$ dimensions is {\it kinematic}, {\it i.e.} whenever the  Lovelock-Ricci vanishes so does the corresponding Riemann. That is, pure Lovelock vacuum in critical odd dimension is Lovelock flat, as is the case for $N=1$ Einstein gravity in $3$ dimension. Based on this result, Dadhich \cite{dad2} conjectured that this should be true not only for static vacuum solutions but in general for all vacuum spacetimes. It would be a universal gravitational property. However it later turned out that this is not true in general for Dadhich's Lovelock-Riemann tensor while it is true for Kastor's analogue \cite{kast}. This is a a purely algebraic property due to the fact that we can write Kastor's $4N$th rank tensor $\mathbb{R}$ (therefore also all its contractions) in terms of the Lovelock-Ricci (or equivalently the corresponding Einstein, $\mathcal{E}^{a}_{\ b}$). As we will describe below, in $d=2N+1$ we can write
\begin{equation}
\mathbb{R}^{b_1\cdots b_{2N}}_{a_1\cdots a_{2N}}=\frac{1}{(2N)!}\epsilon^{b_1\cdots b_{2N+1}}\,\epsilon_{a_1\cdots a_{2N+1}}\mathcal{E}^{a_{2N+1}}_{\qquad b_{2N+1}}~.
\label{main}
\end{equation}
Notice that this also fixes completely the form of $\mathbb{R}$ in presence of a cosmological constant. Drawing again parallels with three dimensional general relativity we will show that there is also a higher order analogue of the Weyl tensor  that vanishes in that case.\\ 

The paper is organized as follows: we will first review Kastor's formulation that can be suitably described in the language of differential forms. This will be particularly convenient for the derivation of the Bianchi identities for the new higher order tensors, and also to find a much more direct route to show the kinematicity of pure Lovelock gravity in odd critical dimensions. Next we reconcile the two formulations showing their equivalence and we end with a discussion.

\section{Kastor's formulation} 

The starting point of Kastor's construction \cite{kast} is a $(2N,2N)$-rank tensor product of $N$ Riemann tensors, completely antisymmetric, both in its upper and lower indices,
\begin{equation}
\left.\right._{(N)}\!\mathbb{R}^{b_1 b_2 \cdots b_{2N}}_{a_1 a_2 \cdots a_{2N}}= R^{[b_1 b_2}_{\quad \quad [a_1 a_2}\cdots R^{b_{2N-1} b_{2N}]}_{\qquad \qquad a_{2N-1} a_{2N}]}~.
\label{kastensor}
\end{equation}
With all indices lowered, this tensor is also symmetric under the exchange of both groups of indices, $a_i\leftrightarrow b_i$. In a similar way we will denote the contractions of $\mathbb{R}$ simply as
\begin{equation}
\left.\right._{(N)}\!\mathbb{R}^{b_1 b_2 \cdots b_{J}}_{a_1 a_2 \cdots a_{J}}=\left.\right._{(N)}\!\mathbb{R}^{b_1 b_2 \cdots b_{J} c_{J+1} \cdots c_{2N}}_{a_1 a_2 \cdots a_{J} c_{J+1} \cdots c_{2N}} \qquad ; \quad \forall \, J<2N ~.
\end{equation}
We will use latin indices generally. When the difference between tangent space and coordinate frames is needed, the latter indices will be denoted with greek letters. In what follows we will omit the index $(N)$ indicating the  Lovelock or curvature order except when not clear from the context.

Rather than working with tensors, in some cases it is convenient to use the language of differential forms and the exterior algebra. This is particularly useful for Lovelock gravities as it makes the expressions much more compact and simplifies a lot many manipulations. Recalling the relation between the curvature  $2$-form and the Riemann tensor $$R^{ab}=\frac12 R^{ab}_{\ \ \mu\nu}\,dx^\mu dx^\nu$$ we can use \reef{kastensor} as components of a $2N$ form which is the antisymmetrized wedge product of $N$ curvature 2-forms,
\begin{equation}
\mathbb{R}^{b_1 b_2 \cdots b_{2K}}_{(N)}= R^{[b_1 b_2}\wedge \cdots \wedge R^{b_{2N-1} b_{2N}]} ~.
\end{equation}
From this $2N$-form it is trivial to construct both the  Lovelock action and its corresponding equation of motion. We just need to complete a $d$-form with vielbeins and contract with  the antisymmetric symbol
\begin{eqnarray}
\mathcal{L}&=&\frac{2^N}{(2N)!(d-2N)!}\epsilon_{a_1 a_2\cdots a_d}\,\mathbb{R}^{a_1a_2\cdots a_{2N}}\wedge e^{a_{2N+1}}\wedge \cdots \wedge e^{a_d} ~,\\
\left.\right.\mathcal{E}_{\ c}^{b}&=& \frac{2^N}{(2N)!(d-2N-1)!}\,\epsilon_{a_1 a_2\cdots a_{d-1} c}\,\mathbb{R}^{a_1a_2\cdots a_{2N}}\wedge e^{a_{2N+1}}\wedge \cdots \wedge e^{a_{d-1}}\wedge e^b ~.
\label{einstein}
\end{eqnarray}
Notice that the antisymmetry properties of the upper indices of $\mathbb{R}$ are completely irrelevant for constructing the action or the equation of motion. We could have defined an analogous tensor without antisymmetrizing the upper indices and the above formulae would have remained unchanged. However all extra terms introduced are irrelevant as they are zero on  contraction with the antisymmetric symbol. Thus 
$\mathbb{R}$  
tensor encodes all the relevant dynamical information with the minimal number of independent components.

Expressions and derivation of Bianchi identities are also simpler in differential form language. This will also be true for the Bianchi identities associated with these new tensors that, due to their high degree of symmetry, have a very simple  form. Let us first step back a bit and explain how Bianchi identities arise in the case of Einstein-Hilbert gravity. In differential form language, the torsion and curvature forms are introduced via Cartan's structure equations,
\vspace{-4mm}
\begin{eqnarray}
T^a &=& De^a=de^a+\omega^a_{\ b}\wedge e^b ~,\label{streq}\\
R^a_{\ b}&=&d\omega^a_{\ b}+\omega^a_{\ c}\wedge \omega^c_{\ b} ~,\nonumber
\end{eqnarray}
for which we have introduced a covariant exterior derivative, $D$, with the corresponding connection 1-form $\omega^a_{\ b}$, in addition to the usual exterior operator $d$.

From Eq.\,\reef{streq} it is easy to derive the corresponding Bianchi identities just using the nilpotency of the exterior derivative, \ie $d^2\!=\!0$ identically. Notice however that the covariant derivative $D$ is not nilpotent. The Bianchi identities can be written simply
\begin{eqnarray}
DT^a&=&R^a_{\ b}\wedge e^b ~,\\
DR^{ab}&=&0 ~.\nonumber
\end{eqnarray}
For vanishing torsion, expressing the above in components we recover the well known expressions
\vspace{-5mm}
\begin{eqnarray}
R^a_{\ [bcd]}&=&0 ~,\\
R^{ab}_{\ \ [\mu\nu;\alpha]}&=&0~. \nonumber
\end{eqnarray}

These expressions have a very easy generalization for $N$th order Lovelock gravity. In the same way as for the curvature 2-form, we can take the exterior covariant derivative of $\mathbb{R}$ and write
\begin{equation}
D\mathbb{R}^{a_1 a_2\cdots a_{2N}}=N\, DR^{[a_1 a_2}\wedge R^{a_3 a_4}\wedge \cdots \wedge R^{a_{2K-1} a_{2K}]}=0
\end{equation}
or again in components
\begin{equation}
\mathbb{R}_{[\mu_1 \mu_2\cdots \mu_{2N};\nu]}^{a_1 a_2\cdots a_{2N}}=0 ~.
\end{equation}
On taking the trace of this identity we get  $\mathcal{E}^a_{\ b;a}=0$ from where we can then extract the required divergence free Lovelock-Einstein tensor \reef{einstein}. In terms of contractions of $\mathbb{R}$ it can be written as 
\begin{equation}
\mathcal{E}^a_{\ b}=-(2N+1)\delta^{aa_1\cdots a_{2N}}_{b b1\cdots b_{2N}}\mathbb{R}^{b_1\cdots b_{2N}}_{a_1\cdots a_{2N}}=2N\mathbb{R}^a_{\ b}-\delta^a_{\ b}\mathbb{R} ~.
\end{equation}
\vskip.5em

In order to obtain the other Bianchi identity we need to contract $\mathbb{R}$ with a vielbein to get 
\begin{equation}
\mathbb{R}^{a_1 a_2\cdots a_{2N}}\wedge e_{a_1}=0
\end{equation}
or equivalently in components,
\begin{equation}
\mathbb{R}_{[a_{2N} b_1 b_2\cdots b_{2N}]}^{a_1 a_2\cdots a_{2N-1}}=0 ~.
\end{equation}
These generalized Bianchi identities trivially reduce to the usual ones for $N=1$. \\

In deriving the form of $\mathcal{E}^a_{\ b}$ we have made use of a couple of very handy identities. The first one allows us to rewrite a contraction of antisymmetric symbols in terms of antisymmetrized products of $\delta$-functions,
\begin{equation}
\epsilon^{a_1\cdots a_kc_{k+1}\cdots c_d}\,\epsilon_{b_1\cdots b_kc_{k+1}\cdots c_d}=-k!(d-k)!\delta^{a_1\cdots a_k}_{b_1\cdots b_k}~,
\label{ident1}
\end{equation}
where 
\begin{equation}
\delta^{a_1 a_2\cdots a_n}_{b_1 b_2\cdots b_n}=\delta^{a_1}_{[b_1}\delta^{a_2}_{b_2}\cdots \delta^{a_n}_{b_n]}=\delta^{[a_1}_{b_1}\delta^{a_2}_{b_2}\cdots \delta^{a_n]}_{b_n}~.
\end{equation}
The second expression involves contractions of the latter
\begin{equation}
\delta^{aa_1\cdots a_{2N}}_{b b_1\cdots b_{2N}}=\frac{1}{2N+1}\left(\delta^a_b \delta^{a_1\cdots a_{2N}}_{b_1\cdots b_{2N}}-2N\delta^a_{[b_1}\delta^{a_1\cdots a_{2N}}_{b b_2\cdots b_{2N}]}\right)~.
\label{ident2}
\end{equation}
This formulae will prove extremely useful to carry out the computations contained in the rest of this note.\\

\section{Kinematicity relative to the Lovelock-Riemann tensor}

Three dimensional Einstein gravity is {\it kinematic}, it does not have local degrees of freedom. This is due to the fact that in three dimensions the Riemann tensor is completely fixed by the Ricci and vice versa. This can already be seen at the level of the number of independent components, both having six, and we can explicitly write
\begin{equation}
R^{ab}_{\ \ cd}=4\delta^{[a}_{[c}R^{b]}_{\ d]}-\delta^{[a}_{[c}\delta^{b]}_{d]}R ~.
\label{zeroweyl}
\end{equation}
Equivalently we can just say that the Weyl tensor identically vanishes in three dimensions, this being also true in the presence of cosmological constant. The equation of motion in vacuum thus fixes completely the form of the Riemann curvature that in turn fixes the metric up to diffeomorphism. The metric itself has to be that of a maximally symmetric space and no local degrees of freedom can propagate. 

Higher order analogues of Eq.\,\reef{zeroweyl} have been derived  in \cite{kast}. There, a new set of dimensional dependent identities has been used to write the $4N$th rank tensor $\mathbb{R}$ in terms of its contractions in dimensions $d<4N$.
The way these identities appear is very easy to understand.
Notice that the expression of $\mathbb{R}$ in Eq.\,\reef{kastensor} is antisymmetrized over sets of $2N$ indices. Thus this tensor vanishes for dimensions $d<2N$ and the corresponding Lovelock term becomes trivial. We can now antisymmetrize over bigger sets of indices respecting the symmetry properties of the above tensor, {\it i.e.} we may define
\begin{eqnarray}
\mathbb{A}^{b_1 b_2 \cdots b_{2N}}_{a_1 a_2 \cdots a_{2N}}&=&\delta^{b_1 b_2\cdots b_{2N} c_1 c_2\cdots c_{2N}}_{a_1 a_2\cdots a_{2N} d_1 d_2\cdots d_{2N}}\, R^{d_1 d_2}_{\quad \ \ c_1 c_2}\cdots R^{d_{2N-1} d_{2N}}_{\qquad \quad c_{2N-1} c_{2N}}\\[.4em]
&=& \delta^{[b_1}_{[a_1}\delta^{b_2}_{a_2}\cdots
\delta^{b_{2N}}_{a_{2N}}\mathbb{R}^{c_1\cdots c_{2N}]}_{c_1\cdots c_{2N}]}\nonumber
\end{eqnarray}
antisymmetrizing over sets of $4N$ indices. This new tensor, that can be written explicitly in terms of $\mathbb{R}$ and its contractions, vanishes for dimensions  $d<4N$. Interestingly, we get a way of writing $\mathbb{R}$ completely in terms of its contractions below that dimensionality. Further, we can define similar tensors reducing the number of free indices on each set
\begin{equation}
\mathbb{A}^{b_1 b_2 \cdots b_{2N-J}}_{a_1 a_2 \cdots a_{2N-J}}=\delta^{[b_1}_{[a_1}\delta^{b_2}_{a_2}\cdots
\delta^{b_{J}}_{a_{J}}\mathbb{R}^{c_1\cdots c_{2N}]}_{c_1\cdots c_{2N}]} \qquad ;\quad \forall\, 0<J<2N~.
\end{equation}
This tensor will vanish for $d<4N-J$ and will allow us to write $J$th contraction of $\mathbb{R}$ in terms of lower contractions below that critical dimension. These identities were used by Kastor \cite{kast} in order to prove that in all odd $d=2N+1$, the $4N$th rank tensor $\mathbb{R}$ can be written completely in terms of its corresponding Lovelock-Ricci, and this is why whenever the latter vanishes so does the former. This is however not the case in next even $d=2N+2$ dimension or higher. \\

The above construction based on identities looks quite simple, however it may become very cumbersome when it comes to derive explicit expressions. In $d=2N+1$ dimensions, for example, in writing explicitly the $4N$th rank tensor in terms of the corresponding Lovelock-Ricci, one has to write down the explicit expressions of the $2N-1$ identities available  for that dimensionality and combine them to get the desired expression. In the next few paragraphs we will describe a much more efficient way of deriving these expressions based on the use of the Hodge duality. In particular, as stated in the introduction, this will lead us to our Eq.\,\reef{main}.

Our starting point is the basic observation that antisymmetric tensors of ranks $n$ and $(d-n)$ have the same number of independent components. In fact there is a reversible map between the two equivalent representations, namely the Hodge duality, that basically amounts to a contraction with the antisymmetric symbol,
\begin{equation}
(\star T)_{a_{n+1}\cdots a_{d}}=\frac{1}{n!}\epsilon_{a_1a_2\cdots a_{n}a_{n+1}\cdots a_d}T^{a_1a_2\cdots a_{n}}~.
\end{equation}
One important advantage of this transformation is that it is easily reversible. Applying it twice we get, up to sign, the original tensor,
\begin{equation}
(\star\!\star\! T)^{a_1\cdots a_n}=\frac{1}{n!(d-n)!}T^{b_1\cdots b_n}\epsilon_{b_1\cdots b_d}\epsilon^{b_{n-1}\cdots b_d a_1\cdots a_n}=(-1)^{1+n(d-n)}\, T^{a_1\cdots a_n} ~.
\end{equation}

In the present case we are also dealing with antisymmetric sets of indices, the only difference being that the tensor of interest has two such sets instead of just one. Either way we can still apply the dual map to each set separately and get an equivalent $(d-2N,d-2N)$-tensor as
\begin{equation}
(\star\mathbb{R}\star)^{b_1 b_2 \cdots b_{d-2N}}_{a_1 a_2 \cdots a_{d-2N}}=\left(\frac{1}{(2N)!}\right)^2\,\epsilon^{b_1 b_2\cdots b_{d-2N} c_1 c_2\cdots c_{2N}}\,\epsilon_{a_1 a_2\cdots a_{d-2N} d_1 d_2\cdots d_{2N}}\, \mathbb{R}^{d_1 d_2 \cdots d_{2N}}_{c_1 c_2 \cdots c_{2N}}
\label{dualR}
\end{equation}
such that $(\star\!\star\! \mathbb{R} \!\star\!\star)=\mathbb{R}$. For $d<4N$ the new tensor will be of lower rank  as compared to the original $\mathbb{R}$. In fact $(\star\mathbb{R}\star)$ will be given by a particular combination of contractions of $\mathbb{R}$ in that case. Applying the Hodge star again we recover the original tensor $\mathbb{R}=\star(\star\mathbb{R}\star)\star$ now expressed in terms of its contractions. This is precisely the identity we were looking for. In particular we can recover all of Kastor's identities from this single one.

Notice that the above Eq.\,\reef{dualR} is very similar in form to the Lovelock-Einstein tensor \reef{einstein}. In the critical dimension $d=2N+1$ we get in fact
\begin{equation}
(\star\mathbb{R}\star)^{b}_{a}=\frac{1}{(2N)!}\,\mathcal{E}^b_{\ a}=\frac{1}{(2N)!}\left(2N\mathbb{R}^a_{\ b}-\delta^a_{\ b}\mathbb{R}\right)
\end{equation}
and vice versa,
\begin{equation}
\mathbb{R}^{b_1\cdots b_{2N}}_{a_1\cdots a_{2N}}=\frac{1}{(2N)!}\epsilon^{b_1\cdots b_{2N+1}}\,\epsilon_{a_1\cdots a_{2N+1}}\mathcal{E}^{a_{2N+1}}_{\qquad b_{2N+1}} ~,
\label{dualrep}
\end{equation}
making explicit what we wanted to prove. When the Lovelock-Ricci tensor vanishes (or equivalently $\mathcal{E}^a_{\ b}$) the whole tensor $\mathbb{R}$ vanishes as well, along with all its contractions. 

In dimensions above the critical one this does not directly apply. In particular, in $d=2N+2$ dimensions in order for the tensor $\mathbb{R}$ to be identically zero, not just the corresponding Ricci has to vanish but also the Lovelock-Riemann has to be zero. This can be easily guessed as $(\star\mathbb{R}\star)$ is a (2,2)-tensor in this case. More explicitly,
\begin{equation}
(\star\mathbb{R}\star)^{ab}_{cd}=\frac{-1}{(2N)!}\left[2\delta^{[a}_{[c}\delta^{b]}_{d]}\mathbb{R}- 8N\delta^{[a}_{[c}\mathbb{R}^{b]}_{d]}
+2N(2N-1)\mathbb{R}^{ab}_{cd}\right] ~,
\end{equation}
and the whole tensor $\mathbb{R}$ can be written in terms of its Lovelock-Riemann and its contractions in this case. The higher we go in dimension, the higher the rank of the contractions involved in the  $(\star\mathbb{R}\star)$ and $\mathbb{R}$ expressions. This reflects the fact that, in Kastor's approach, we have less identities to play with.

As a check of our formulae we can compute the Lovelock-Riemann tensor in terms of the double dual tensor,
\begin{equation}
\mathbb{R}^{ab}_{cd}=-(2N-2)!\left[\delta^{[a}_{[c}\delta^{b]}_{d]}(\star\mathbb{R}\star)-4\delta^{[a}_{[c}(\star\mathbb{R}\star)^{b]}_{d]}+(\star\mathbb{R}\star)^{ab}_{cd}\right]
\end{equation}
then, plugging the explicit expression of $(\star\mathbb{R}\star)$ and its contractions we can see that in fact the right hand side yields $\mathbb{R}^{ab}_{cd}$. We can also check that the Lovelock-Einstein tensor can be written as
\begin{equation}
(\star\mathbb{R}\star)^a_c=(\star\mathbb{R}\star)^{ab}_{cb}= \frac{1}{(2N)!}\mathcal{E}^a_{\ c} ~,
\end{equation}
a contraction of $(\star\mathbb{R}\star)$ in $d=2N+2$.\\

The previous discussion can be trivially modified to include a nonzero cosmological constant, the above equations being still valid in that case. We just have to modify the equation of motion to
\begin{equation}
\mathcal{E}^a_{\ b}=\lambda \delta^a_b ~,
\end{equation}
such that, instead of zero, the Lovelock-Ricci is now proportional to the metric. In odd critical dimensions we can again use Eq.\,\reef{dualrep} and verify that the form of the tensor $\mathbb{R}$ is completely fixed to
\begin{equation}
\mathbb{R}^{a_1\cdots a_{2N}}_{b_1\cdots b_{2N}}=\lambda \delta^{a_1\cdots a_{2N}}_{b_1\cdots b_{2N}} ~.
\end{equation}
Analogously to what happens for Einstein gravity we can define a Lovelock-Weyl 
\begin{equation}
\mathbb{W}^{ab}_{cd}=\mathbb{R}^{ab}_{cd}-\frac{4}{d-2}\delta^{[a}_{[c}\mathbb{R}^{b]}_{d]}+\frac{2}{(d-1)(d-2)}\delta^{[a}_{[c}\delta^{b]}_{d]}\mathbb{R}
\end{equation}
that then vanishes in odd $d=2N+1$ dimensions whereas it is completely unconstrained in $d=2N+2$ or higher. In \cite{confkast} other higher order analogues of the Weyl tensor have been constructed that vanish for $d<4N$.
\\

To sum up, in any dimension we can always write $\mathbb{R}=(\star\!\star\!\mathbb{R}\!\star\!\star)$ and thus express $\mathbb{R}$ in terms of $(\star\mathbb{R}\star)$. For low enough dimension, $d<4N$, the dual tensor itself will be a combination of contractions of $\mathbb{R}$ with the same number of free indices. In $d=2N$, $(\star\mathbb{R}\star)\sim\mathcal{L}$ is just a scalar what means that $\mathbb{R}$ can be written solely in terms of the Lovelock scalar, in $d=2N+1$ it can be written in terms of the Lovelock-Ricci and its trace, in $d=2N+2$ we need to include also the Lovelock-Riemann and so on.  

The above discussion implies that for pure Lovelock gravity in the critical dimension the vacuum equation of motion (with or without cosmological constant) $\mathcal{E}^a_{\ b}=\lambda \delta^a_b$ completely fixes $\mathbb{R}$ and all its contractions. Thus we have proved in a very direct way that Lovelock gravity is {\it kinematic} in all odd $d=2N+1$ dimensions as it has no degrees of freedom relative to the Lovelock-Riemann tensor. However, unlike in $d=3$ this does not fix completely the Riemann tensor, thus our solutions are not locally maximally symmetric spaces and we have in general propagating degrees of freedom. For zero $\lambda$ we can rephrase this by saying that any solution of pure Lovelock is {\it Lovelock flat} even though the Riemann curvature is not necessarily zero. For nonzero $\lambda$ the Lovelock-Weyl tensor is zero but still the Weyl tensor does not necessarily vanish. The implications of this for the dynamics of pure Lovelock theories is still unclear.

\section{Kastor - Dadhich reconciliation}

As we have described in the previous section, Kastor's tensors contain all the relevant information from which we can reconstruct action and equation of motion in any pure Lovelock theory. In this way, any other formulation that cannot be obtained from this one would contain more information on the spacetime that does not enter either the action or the corresponding Lovelock-Einstein tensor. In particular, Kastor's $4N$th rank tensor is totally antisymmetric on each set of $2N$ indices and symmetric under exchange of both sets. In addition the dualization procedure allowed us to write $\mathbb{R}$ in terms of its contractions in dimensions $d<4N$. This reduces the information contained in $\mathbb{R}$ to the minimal amount still capturing the whole dynamics. Any extra information is thus irrelevant from this point of view. \\

Dadhich \cite{dad1} proposed a different set of tensors as a suitable tool for describing the dynamics of pure Lovelock gravity. For it to be equivalent to Kastor's description, it should be possible to write everything in terms of $\mathbb{R}$. We will see that this is actually not possible and that we will need to add a new tensor structure to the ones used by Dadhich in order to recover Kastor's Lovelock-Riemann. The basic object in Dadhich's formulation is a $4$th rank Riemann analogue tensor which is a $N$th order homogeneous polynomial in the Riemann curvature. It is given by  
\begin{equation}
\mathcal{R}_{abcd}=F_{abcd}-\frac{N-1}{N(d-1)(d-2)}F\left(g_{ac}g_{bd}-g_{ad}g_{bc}\right)
\label{nartensor}
\end{equation}
where
\begin{eqnarray}
F^{cd}_{\ \ ab}&=&\left(\delta_{abc_1d_1\cdots c_{N-1}d_{N-1}}^{mna_1b_1\cdots a_{N-1}b_{N-1}}R^{c_1d_1}_{\quad\ a_1b_1}\cdots R^{c_{N-1}d_{N-1}}_{\qquad a_{N-1}b_{N-1}}\right)R^{cd}_{\ \ mn}\\
&=& R^{c_1d_1}_{\ \ [c_1d_1}\cdots R^{c_{N-1}d_{N-1}}_{\qquad c_{N-1}d_{N-1}}R^{cd}_{\ \ ab]}= R^{[c_1d_1}_{\quad [c_1d_1}\cdots R^{c_{N-1}d_{N-1}]}_{\qquad c_{N-1}d_{N-1}}R^{cd}_{\ \ ab]}\nonumber ~.
\end{eqnarray}
Written in this way the tensor $F$ looks very similar to the Riemann contraction of $\mathbb{R}$, except that upper indices are not antisymmetrized. The contracted indices can be considered as antisymmetrized as lower indices are, but not the whole set. We shall now extract the difference between the two classes of tensors. Comparing the above $F$ with the Lovelock-Riemann in Kastor's formulation,
\begin{equation}
\mathbb{R}^{cd}_{ab} = R^{[c_1d_1}_{\quad\ [c_1d_1}\cdots R^{c_{N-1}d_{N-1}}_{\qquad c_{N-1}d_{N-1}}R^{cd]}_{\ \ ab]} ~,
\end{equation}
we can easily see that $F$ is not symmetric under the exchange of both pairs of indices (when all lowered) whereas this contraction of $\mathbb{R}$ is. By repeatedly using Eq.\,\reef{ident2}, $\mathbb{R}^{cd}_{ab}$ can be rewritten  as
\begin{eqnarray}
\mathbb{R}^{cd}_{ab}&=&\frac{1}{N(2N-1)}\left(\left[1+(N-1)(2N-3)\right]F^{cd}_{\ \ ab}\right.\\[.4em]
&-& \left. 4(N-1)\,R^{ca_1}_{\quad [ab}R^{db_1}_{\quad a_1 b_1}\cdots R^{a_{N-1}b_{N-1}}_{\qquad a_{N-1}b_{N-1}]}\right) ~.\nonumber
\end{eqnarray}
The difference between $\mathbb{R}$ and $F$ is the second term in the bracket, that is the only other structure that can be written respecting all the relevant symmetries. Both tensors are equal in the trivial case of $N=1$. Note that Lovelock-Ricci tensors arising from $F$ and $\mathbb{R}$ are the same
\begin{equation}
\mathbb{R}^a_{\, b}=F^a_{\;\ b}
\label{2ricci}
\end{equation}
but this does not mean that the extra contribution is traceless. Instead we can write
\begin{equation}
R^{ca_1}_{\quad [ab}R^{bb_1}_{\quad a_1 b_1}\cdots R^{a_{N-1}b_{N-1}}_{\qquad a_{N-1}b_{N-1}]} = - \mathbb{R}^{cb}_{ab}=- \mathbb{R}^{c}_{a}
\end{equation}
which clearly leads to \reef{2ricci}. The contractions of the other  4th rank tensor $\mathcal{R}_{abcd}$ are different though because of the scalar piece in \reef{nartensor},
\begin{equation}
\mathcal{R}^a_{\ b}=\mathbb{R}^a_{\ b}-\frac{N-1}{N(d-2)}\mathbb{R}\delta^a_{\ b}~,
\end{equation}
\begin{equation}
\mathcal{R}=
\frac{d-2N}{N(d-2)}\mathbb{R}~.
\end{equation}
This difference pops up also in the corresponding Lovelock-Einstein tensors that have different normalizations for their Ricci and scalar pieces, 
\begin{equation}
\mathcal{E}^a_{\ b}=2N\left(\mathcal{R}^a_{\ b}-\frac12 \mathcal{R}\delta^a_{\ b}\right)
=2N\mathbb{R}^a_{\ b}- \mathbb{R}\delta^a_{\ b}
\end{equation}
Upon contraction we get
\begin{equation}
\mathcal{E}^a_{\ a}=-N(d-2)\mathcal{R}
=-(d-2N)\mathbb{R}
\end{equation}
showing that the scalars in both formulations also have different normalizations, something that has to be taken into account at the level of the action. Kastor's parametrisation makes explicit the fact that in $d=2N$ the Lovelock action (or $\mathbb{R}$) is a topological density and its variation is therefore zero, $\mathcal{E}^a_{\ a}=0$. In Dadhich's  case, this fact gets obscured by the fact that his scalar $\mathcal{R}$ includes an extra $(d-2N)$ factor and thus vanishes as well for $d=2N$.

\subsection{Pure Gauss-Bonnet check}

As a check of the above expressions we will analyze the $N=2$ case of pure Gauss-Bonnet (GB) gravity for which we may compute
\begin{equation}
\left.\right.\mathbb{R}^{cd}_{\ \ ab}=\frac{1}{3}\left(\left.\right.F^{cd}_{\ \ ab}-2R^{ca_1}_{\quad [ab}R^{db_1}_{a_1 b_1]}\right)
\label{Rgb}
\end{equation}
and 
\begin{equation}
\left.\right.F^{cd}_{\ \ ab}=\frac16\left(RR^{cd}_{\ \ ab}+4R^{cd}_{\ \ k[a}R^k_{\ b]}+R^{cd}_{\ \ kl}R^{kl}_{ \ \ ab}\right). 
\end{equation}
Let us also write the second term explicitly as 
\begin{equation}
2R^{ca_1}_{\quad [ab}R^{d b_1}_{a_1 b_1]}=\frac23\left(R^{[c|k|}_{\quad ab}R^{d]}_{\ \ k}-R^{[c}_{\ \ [a}R^{d]}_{\ \ b]}+R^{k[c}_{\quad l[a}R^{d]l}_{\quad b]k}\right) ~ .
\end{equation}

As we have seen in previous sections, in $d=2N+1$ dimensions the higher rank tensor $\mathbb{R}$ vanishes for vacuum solutions of pure Lovelock gravity with zero cosmological constant. This tensor is the double Hodge dual of the Lovelock-Einstein tensor (see Eq.\,\reef{dualrep}). It vanishes whenever $\mathcal{E}^a_{\ b}=0$ and so does  any of its contractions, namely Kastor's Lovelock-Riemann tensor. Moreover, we know that both formulations agree at the Ricci tensor level. Thus, whenever the Lovelock-Ricci vanishes in $d=2N+1$ dimensions, $F$ and the second term in Eq.\,\reef{Rgb}
exactly cancel out each other. For spherically symmetric solutions it turns out that $F$ and the extra term vanish separately in $d=2N+1$, and that is the basis for Dadhich's conjecture \cite{dgj, dad2} for the kinematicity of pure Lovelock gravity in odd critical dimensions. As an example for which both contributions to the Lovelock-Riemann are separately nonzero we can consider a pure GB Kasner vacuum in five dimensions \cite{alfred}. One such metric is, for instance,
\begin{equation}
ds^2=-dt^2+t^{2p_1 }dx_1^2+t^{2p_2}dx_2^2+t^{2(3-p_1-p_2) }dx_3^2+dx_4^2 ~,
\end{equation}
where $p_{1,2}$ are arbitrary constants. In this case the nonzero $F$ is cancelled out by the extra term to give $\mathbb{R}=0$.

\section{Discussion} 

It is interesting that there are two parallel but distinct definitions of a higher order Lovelock-Riemann tensor leading to the same equation of motion. That is, the two constructions describe precisely the same gravitational dynamics, even though there is a nontrivial difference at a kinematic level. This became apparent when a pure GB Kasner vacuum solution was found in five dimensions \cite{alfred} for which Dadhich's Lovelock-Riemann tensor did not vanish. Besides this was in contradiction with a previous kinematicity conjecture \cite{dgj,dad2}. Dadhich's tensor did indeed vanish for spherically symmetric pure GB vacuum solutions \cite{dgj}, and based on that it was proposed that any pure Lovelock vacuum in all odd $d=2N+1$ dimensions was Lovelock flat - {\it kinematic}. A precise realization of the kinematicity property was nonetheless provided by an  alternative formulation put forward by Kastor \cite{kast}. Kastor's Lovelock-Riemann tensor did vanish for the Kasner vacuum solution in question. In fact this tensor vanishes in all odd critical dimensions for {\it any} vacuum solution of pure Lovelock gravity. It became thus pertinent to reconcile both formulations, and this is therefore the main motivation of this investigation. Both Lovelock-Riemann tensors differ in a piece that, remarkably, vanishes in the spherically symmetric case and that is how it was not noticed at first \cite{dgj}. \\

We have also revisited this kinematicity property and proved it in a much more direct way making extensive use of the properties of the Hodge dual map. This is a unique and universal distinguishing feature of pure Lovelock gravity in all odd $d=2N+1$ dimensions which is shared by no other theory. It stems from the fact that, for that critical dimensionality, we can write the higher rank tensor $\mathbb{R}$ as the double Hodge dual of the corresponding Lovelock-Einstein tensor. Thus $\mathbb{R}$ is completely fixed by the equations of motion. It is important to note that this {\it kinematicity} is relative to the Lovelock-Riemann tensor and not to the Riemann curvature. That is, the Lovelock-Riemann tensor vanishes in $d=2N+1$ whenever the corresponding Lovelock-Ricci vanishes, but the Riemann curvature may be nonzero. In turn, in $d=2N+2$, the Lovelock-Weyl tensor is {\it a priori} unrelated to the equation of motion. This is in complete analogy with the behavior of Einstein gravity in three and four dimensions and it can be also generalized in presence of nonzero cosmological constant. Pure Lovelock gravity thus unravels a new universal feature of gravity in higher dimensions.\\

Pure Lovelock theories posses many interesting properties. Besides the ones already mentioned, thermodynamical parameters of pure Lovelock static black holes bear a universal relation to the horizon radius \cite{dpp} and bound orbits exist in all even $d=2N+2$ dimensions \cite{dgj2} in these spacetimes. It should be pointed out that for Einstein gravity bound orbits around a static black hole exist in $4$ dimensions only. All this strongly suggest that pure Lovelock equation is the right equation to describe the gravitational dynamics in higher $d=2N+1,\, 2N+2$ dimensions \cite{dad2} such that we preserve many interesting properties that Einstein gravity has only for $d=3,4$ respectively. 

\section*{Acknowledgements}

We wish to thank David Kastor for valuable discussions and for first verifying the vanishing of his Riemann tensor for the Kasner solution \cite{alfred} in question. XOC is grateful to Jamia Millia Islamia and IUCAA for hospitality at the initial stages of this work. ND thanks the Albert Einstein Institute, Golm for a visit during which the project was formulated.



\begin{thebibliography}{99}

\bibitem{dad1}
N. Dadhich, ``Characterization of the Lovelock gravity by Bianchi derivative,'' \href{http://dx.doi.org/10.1007/s12043-010-0080-1}{{\em Pramana} {\bf 74} (2010) 875}; \href{http://arxiv.org/abs/arXiv:0802.3034}{{\ttfamily arxiv:0802.3034 [gr-qc]}}.

\bibitem{kast} 
D. Kastor, ``The Riemann-Lovelock curvature tensor,'' \href{http://dx.doi.org/10.1088/0264-9381/29/15/155007}{{\em Class. Quant. Grav.} {\bf 29} (2012) 155007}; \href{http://arxiv.org/abs/1202.5287}{{\ttfamily arxiv:1202.5287 [hep-th]}}.

\bibitem{dgj}
N. Dadhich, S. Ghosh and S. Jhingan, ``The Lovelock gravity in the critical spacetime dimension,'' \href{http://www.sciencedirect.com/science/article/pii/S0370269312003826}{{\em Phys. Lett. B} {\bf 711} (2012) 196}; \href{http://arxiv.org/abs/1202.4575}{{\ttfamily arxiv:1202.4575 [gr-qc]}}.

\bibitem{dad2} 
N. Dadhich, ``The gravitational equation in higher dimensions,'' in \href{http://dx.doi.org/10.1007/978-3-319-06761-2_6}{{\em Relativity and Gravitation, 100 years after Einstein in Prague}, eds J. Bicak and T. Ledvinka, Springer (2013)}; \href{http://arxiv.org/abs/arXiv:1210.3022}{{\ttfamily arxiv:1210.3022 [gr-qc]}}.

\bibitem{confkast}
D. Kastor, ``Conformal Tensors via Lovelock Gravity,'' \href{http://dx.doi.org/10.1088/0264-9381/30/19/195006}{{\em Class. Quant. Grav.} {\bf 30} (2013) 195006}; \href{http://arxiv.org/abs/arXiv:1306.4637}{{\ttfamily arxiv:1306.4637 [gr-qc]}}.

\bibitem{alfred}
N. Dadhich, A. Molina and X. O. Camanho, ``Pure Lovelock Kasner metric,'' to appear. 

\bibitem{dpp}
N. Dadhich, J. M. Pons, K. Prabhu, ``Thermodynamical universality of the Lovelock black holes,'' \href{http://dx.doi.org/10.1007/s10714-012-1416-6}{{\em Gen. Rel. Grav.} {\bf 44} (2012) 2595}; \href{http://arxiv.org/abs/arXiv:1110.0673}{{\ttfamily arXiv:1110.0673 [gr-qc]}}.

\bibitem{dgj2}
N. Dadhich, S. G. Ghosh, S. Jhingan, ``Bound orbits and gravitational theory,'' \href{http://dx.doi.org/10.1103/PhysRevD.88.124040}{{\em Phys. Rev. D} {\bf 88} (2013) 124040}; \href{http://arxiv.org/abs/arXiv:1308.4770}{{\ttfamily arXiv:1308.4770 [gr-qc]}}.



\end{thebibliography}
\end{document}